\documentclass[dcolumn]{revtex4}    

\usepackage{dcolumn}
\usepackage{graphicx}
\usepackage{mathrsfs}
\usepackage{amsmath}
\usepackage{amsfonts} 

\begin{document}

\title{Prediction of temperature distribution in turbulent Rayleigh-Benard convection}
\author{Zhen-Su She} \email{she@pku.edu.cn}
\affiliation{State Key Laboratory for Turbulence and Complex Systems
and Department of Mechanics, College of Engineering, Peking
University, Beijing 100871, China}
\author{Xi Chen}
\affiliation{State Key Laboratory for Turbulence and Complex Systems
and Department of Mechanics, College of Engineering, Peking
University, Beijing 100871, China}
\author{Hong-Yue Zou}
\affiliation{State Key Laboratory for Turbulence and Complex Systems
and Department of Mechanics, College of Engineering, Peking
University, Beijing 100871, China} \affiliation{Department of
Mechanics, College of Engineering, Sun Yat-sen University,
Guangzhou, 510275, China}
\author{Yun Bao}
\affiliation{State Key Laboratory for Turbulence and Complex Systems
and Department of Mechanics, College of Engineering, Peking
University, Beijing 100871, China} \affiliation{Department of
Mechanics, College of Engineering, Sun Yat-sen University,
Guangzhou, 510275, China}
\author{Jun Chen}
\affiliation{State Key Laboratory for Turbulence and Complex Systems
and Department of Mechanics, College of Engineering, Peking
University, Beijing 100871, China}
\author{Fazle Hussain}
\affiliation{State Key Laboratory for Turbulence and Complex Systems
and Department of Mechanics, College of Engineering, Peking
University, Beijing 100871, China} \affiliation{Department of
Mechanical Engineering, University of Houston, Houston, TX
77204-4006, USA}


\date{\today}

\begin{abstract}

A quantitative theory is developed for the vertical mean temperature
profile (MTP) in turbulent Rayleigh-Benard convection (RBC), which
explains the recent experimental and numerical observations of a
logarithmic law by Ahlers et al. \cite{Ahlers2012a}. A multi-layer
model is formulated and quantified, whose predictions agree with DNS
and experimental data for the Rayleigh-number ($Ra$) over seven
decades. In particular, a thermal buffer layer follows a 1/7 scaling
like the previously postulated mixing zone \cite{Procaccia1991}, and
yields a $Ra$-dependent log law constant.
A new parameterization of $Nu(Ra)$ dependence is proposed, based on
the present multi-layer quantification of the bulk MTP.
.

\end{abstract}

 \maketitle


Turbulent Rayleigh-Benard convection (RBC) is a well-known
non-equilibrium system displaying rich dynamics of both fundamental
interest and wide technological relevance
\cite{Kadanoff2001,Ahlers2009,DK2010}. Numerous results have been
obtained experimentally and numerically, while the theoretical
understanding remains relatively poor. For instance, a diverse set of heat flux
measurements \cite{He2012a, Ahlers2012b} are yet to be explained;
the origin of observed "classical state", "transition range" and
"ultimate state" \cite{He2012b} remains elusive. Recently, efforts begin to be
directed to the study of the structure of the thermal boundary
layers and of bulk non-vanishing temperature gradient (see, for
instance, \cite{Zhou2010,Stevens2012}), but no accurate prediction of
the full MTP is available. Very recently,
Ahlers et al. \cite{Ahlers2012a} found, beyond the thin boundary
layer unresolved in the experiment, that the temperature and its
root-mean-square (rms) fluctuation vary logarithmically as a
function of the distance $z$ from the bottom plate. Grossman and Lohse
\cite{Gross2011} have developed an argument for the logarithmic
mean temperature with an assumption of kinetic logarithmic layer,
but offer few quantitative predictions. It is clear that a theory of RBC needs to
go beyond the estimates of global quantities \cite{Gross2000} and predict
internal temperature distribution. The recent finding of Alhers et al.
\cite{Ahlers2012a} represents a significant effort in this direction.

Here, we develop a theory of MTP, based on a new mean-field approach
applying a symmetry analysis to wall-bounded turbulent flows
\cite{She10} which yields accurate predictions of the mean velocity
profile over a wide range of Reynolds number \cite{She12}. Here, it
is shown that this mean-field theory yields a valid description of
the MTP in RBC, including, in particular, the log law with
coefficients quantitatively validated by data for $Ra$ over seven
decades. An intriguing outcome is that a thermal buffer layer thickness
is found to vary as $Ra^{1/7}$, exactly like
that of earlier speculated mixing zone \cite{Procaccia1991}.

The temperature variation in fully developed RBC is described by the
Boussinesq equations \cite{Siggia},
\begin{equation}\label{Diff-MTE}
\kappa \frac{{\partial ^2 \Theta }}{{\partial z^2 }} -
\frac{{\partial \left\langle {w\theta } \right\rangle }}{{\partial
z}} = 0
\end{equation}
where $\theta$ denotes temperature re-scaled by the total
temperature difference $\Delta$, and equals $\pm 1/2$ on
the bottom ($z=0$) and top ($z=1$) plates, and $\left\langle ...\right\rangle$ denotes time
and space average except in the vertical $z$ direction
(i.e. $\Theta=\left\langle\theta\right\rangle$); $\kappa$ is the thermal
diffusivity, $w$ is the vertical velocity, carrying all information
about turbulent fluctuations. Integrating (\ref{Diff-MTE}) along $z$ once yields
the mean temperature equation (MTE) \cite{Siggia}:
\begin{equation}\label{MTE}
 - \frac{{d\Theta }}{{dz}} + \frac{{\left\langle {w\theta } \right\rangle }}{\kappa } = \left. { - \frac{{d\Theta }}{{dz}}} \right|_{z = 0}  \equiv Nu
\end{equation}
Denote $ S_\theta ^ +   =  - d\Theta /(Nudz) =  - d\Theta /dz^ +$
and $ W_\theta ^ +   = \left\langle {w\theta } \right\rangle
/(\kappa Nu) =  \left\langle {w^ +  \theta } \right\rangle $ , Then,
the normalized MTE is
\begin{equation}\label{scale-MTE}
S_\theta ^ +   + W_\theta ^ +   = 1
\end{equation}
with normalized vertical coordinate,  $z^+=zNu$ and vertical
velocity, $ w^ +   = w/(\kappa Nu)$.

It is interesting to compare
(\ref{scale-MTE}) to the mean momentum equation (MME) in a
pipe or channel \cite{Lvov2008}: $ \nu dU/dy - \left\langle {uv}
\right\rangle  = u_\tau ^2 (1 - y)$ with friction velocity $u_\tau$,
and kinematic viscosity $\nu$. The MME can be rewritten in wall
units:
\begin{equation}\label{MME}
S^ +   + W^ +   = \frac{{dU^ +  }}{{dy^ +  }} - \left\langle {uv}
\right\rangle ^ +   = 1 - y^ +  /{\mathop{\rm Re}\nolimits} _\tau
\end{equation}
where $ U^ +   = U/u_\tau$, $ W^ +   =  - \left\langle {uv}
\right\rangle ^ +   =  - \left\langle {uv} \right\rangle /u_\tau
^2$, $ y^ +   = yu_\tau  /\nu$ the distance to the wall, $
{\mathop{\rm Re}\nolimits} _\tau = u_\tau  H/\nu$ with $H=1$ being
the half width of a channel. In the so-called overlap region at high
$Re_\tau$, $ y^ +   < < {\mathop{\rm Re}\nolimits} _\tau$, $ S^ + +
W^ + \approx 1$, which is exactly the same as (\ref{scale-MTE}).
Thus, we identify similarities between (\ref{scale-MTE}) and
(\ref{MME}) in three aspects: first, near the wall, both $S^+$ and
$S^+_\theta$ dominate and equal to unity; second, away from the
wall, $W^+$ and $W^+_\theta$ dominate and equal to unity; third,
$W^+$ and $W^+_\theta$ both represent effect of transport by
velocity fluctuations normal to the wall. Thus, we postulate a system
similarity between the momentum and temperature (energy) and suggest
that the distribution of the mean temperature follows the same
symmetry as the mean momentum.

Recently, a symmetry-based theory is developed for $S^+$, giving
rise to a quantitative description of the mean velocity distribution
\cite{She10,She12}, assuming a dilation-group invariance of the
mixing length, $ \ell _M^ += \sqrt {W^ +  } /S^ +$. In analogy to
order parameter displaying a symmetry change during a phase
transition, the mixing length is called order function, which
characterizes the far-from-equilibrium state of turbulent
wall-bounded flow in terms of multiple scaling symmetries occurring
in different domains. Specifically, wall-bounded turbulence forms
several layers, called viscous sublayer, buffer layer, bulk flow
etc., each of which is occupied by fluctuations of distinct scaling
(dilation invariance). The Lie-group analysis is an effective tool
to quantify such a multi-layer structure, yielding an accurate
description from $S^+$-dominant boundary to $W^+$-dominant bulk. The
system similarity assumption guides a similar description for
$\Theta$, by defining a thermal mixing length:
\begin{equation}\label{ltheta}
\ell _\theta ^ +   = \sqrt {W_\theta ^ +  } /S_\theta ^ +
\end{equation}
which is assumed to display the same symmetry property as
$\ell_M^+$.

Note that the dimension of $\ell_\theta$ (the absence of superscript '+'
denotes no normalization) is no longer length, so its
meaning needs to be interpreted. We follow Prandtl's
original apt argument that the effect of a fluctuating velocity on
the transport of a mean "density" (momentum or energy) can be
represented by an eddy viscosity, $ W = \nu _t S$, $W$ being the
Reynolds stress, $ \nu _t  = \ell _M^{2} S$ and $\ell_M$ have the
right dimensions of viscosity and length. In the case
of temperature, $ \nu _\theta   = W_\theta  /S_\theta$ has still the
dimension of diffusivity, but a dimensionally correct expression for
$ \nu _\theta$ should be $ \nu _\theta   = \ell _\theta ^2 S = \ell
_\theta ^2 S_\theta  \left( {S/S_\theta  } \right) = \ell _\theta ^2
S_\theta  \Omega$, where we introduce a dimensional quantity $
\Omega = S/S_\theta$, by substituting $S$ (velocity gradient or
mean shear rate) by $S_\theta$ (temperature gradient). The system
similarity is valid if $ \Omega  \equiv Const$, or $\ell_\theta$
plays the same role for the convective heat transport as $\ell_M$
for momentum transport. This is sound at least when the turbulent
Prandtl Number is unity, because the turbulent velocity and
temperature fluctuations and the corresponding momentum and energy
transports being due to the same fluid motion.

\begin{figure}
\includegraphics[width=7cm]{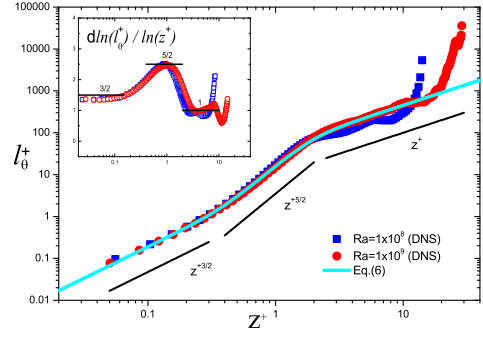}
\caption{(color). The variation of the ``Mixing length order
function" with the normalized vertical distance from the bottom
plate. Symbols are our DNS data at two $Re$s, and line is the model (\ref{lSED}) at
$Ra=10^9$ with $\rho\approx 6$, $z^+_{sub}\approx 0.375$ and $z^+_{buf}\approx 2$.
The inset shows the diagnostic function
$d\ln(\ell_\theta)/d\ln(z)$ displaying the local scaling exponents
of $3/2$, $5/2$ and $1$ for the thermal sublayer, buffer and log layer being, respectively.}
\label{fig:ltheta}
\end{figure}

Whether $\ell_\theta$ is an order function is ultimately tested
against empirical data, by the criterion that it characterizes
multiple scaling states of varying symmetry (or dilation
invariance). The DNS for RBC is performed in a narrow rectangular
box with aspect ratios of X:Y:Z=6:1:6, where $x$, $y$, $z$ denote
horizontal, spanwise and vertical directions, respectively (aspect
ratio $\Gamma=1$). Thermal convection takes place essentially in
the $x$ and $z$ planes, with developed spanwise motions enriching
3-D dynamics. The simulation is performed using a finite-difference
solver of the standard Navier-Stokes equation under Boussinesq
approximation coupled to thermal advection-diffusion equation. Two
simulations are done at $Ra=10^8$ and $10^9$, and $\ell_\theta$ is
computed using (\ref{ltheta}) and shown in Fig.\ref{fig:ltheta}. The
three layers (sublayer, buffer and log layers) are clearly
displayed, while we ignore the last one near the centre for
simplicity, which contribute little at high $Ra$. In this
approximation, the classical laminar thermal boundary layer view of
RBC at this rather moderate $Ra$ is replaced by a three-layer
picture.

Now, we quantify the three-layer temperature variation by
postulating a similar (thermal) sublayer, buffer layer and a log
layer, for which $ \ell _\theta$ displays distinct scalings in z.
Specifically, $\ell _\theta ^ +   \propto \left( {z^ +  }
\right)^{3/2}$ for $z^ + < z_{sub}^ + $; $ \ell _\theta ^ + \propto
\left( {z^ +  } \right)^{5/2}$ for $z_{sub}^ +   < z^ +   < z_{buf}^
+$; and $ \ell _\theta ^ +   \propto z^ +$ for $z^+ > z_{buf}^ +
$. The first exponent ($3/2$) readily follows from a
near-wall expansion with small fluctuations, and the
third linear scaling corresponds to the (postulated) log layer.
However, the thermal buffer layer has a different scaling increment of one, instead of 1/2 for the
momentum buffer-layer, reflecting possibly new form of the near wall
temperature fluctuation structure compared to coherent structures of momentum
fluctuations. While its nature needs further elucidation,
the exponent $5/2$ is empirically confirmed by our DNS data using a Lie-group diagnostic
function \cite{She12}, shown in the inset of Fig.~\ref{fig:ltheta}.
Thus, our system similarity hypothesis yields a composite formula for $\ell_\theta^+$ as:
\begin{equation}\label{lSED}
\ell _\theta ^ +   \approx \rho \left( {z^ +  }
\right)^{\frac{3}{2}} \left( {1 + \left( {\frac{{z^ +  }}{{z_{sub}^
+  }}} \right)^4 } \right)^{\frac{1}{4}} \left( {1 + \left(
{\frac{{z^ +  }}{{z_{buf}^ + }}} \right)^4 } \right)^{ -
\frac{3}{8}}
\end{equation}
where the two transition sharpness parameters are pre-set to be 4
(positive integer), following \cite{She12}. Jointly solving
(\ref{scale-MTE}), (\ref{ltheta}) and (\ref{lSED}) yields a
theoretical MTP as
\textbf{\begin{eqnarray} \frac{1}{2}-\Theta(z^+)
=  \int_0^{z^+} S_\theta dz'= \int_0^{z^+} \frac{{1 - \sqrt {4\ell
_\theta ^{+2} + 1} } }{2\ell _\theta ^{+2} } dz'\label{SED-theta}
\end{eqnarray}}


\begin{figure}
\includegraphics[width=7cm]{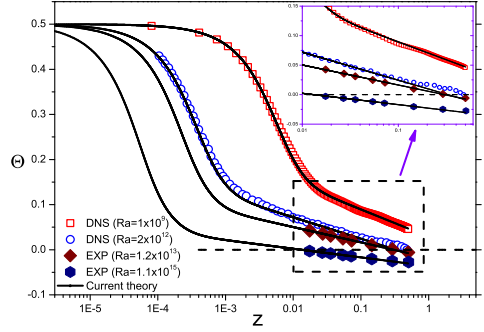}
\caption{(color). Predicted MTP from (\ref{lSED}) and (\ref{SED-theta}) compared to four sets of empirical
data: DNS at moderate-$Ra$ (red squares), DNS at high-$Ra$ (blue
diamonds), and two sets of experimental data at very high $Ra$ (filled
symbols). Data other than the lowest $Ra$ DNS ones are provided by
Ahlers [1]. The inset shows the enlargement
showing agreement of the log profiles in the bulk region.}
\label{fig:Theta}
\end{figure}

The first consequence of the theory is the logarithmic profile.
For $ z^ +  \gg z_{buf}^ +$, $ \ell _\theta ^ +   \approx \kappa
_\theta  z^ +\gg 1$, then $ S_\theta ^ +   \approx 1/\left( {\kappa
_\theta  z^ +  } \right)$, a logarithmic MTP follows:
\begin{eqnarray}
\Theta  &\approx&  - \frac{1}{{\kappa _\theta  }}\ln z^ +   + B^ +
\approx  - A\ln z + B\label{log} \\ \kappa _\theta   &=& \rho
z_{buf}^{ + 3/2} /z_{sub}^ + \label{kapa}
\end{eqnarray}
where coefficients $A$ and $B$ are measured by Ahlers et al.
\cite{Ahlers2012a}. So, $\kappa_\theta$ ($=1/A$) depends on
$Ra$ through three parameters - the central topic investigated below.

The first two parameters specify the thicknesses of the thermal
sublayer and buffer layer, and the third is a global coefficient
determining a Karman-like coefficient. The two thicknesses
characterize the transition between different scaling ranges, and can be
roughly obtained by inspecting Fig.~\ref{fig:ltheta}: $ z_{sub}^ +
\approx 0.35$, $ z_{buf}^+\approx 2$ at our moderate $Ra$. We choose
a fixed $z_{sub}^ +   \approx 0.375$, as expected for an
$Ra$-independent sublayer in thermal boundary units, and then
adjust $z_{buf}^ +$ and $\rho$ for matching the bulk behaviour,
e.g. $A$ and $B$ in (\ref{log}), to yield a prediction of the MTP.
Fig.~\ref{fig:Theta} shows the comparison between the theoretical
profiles with empirical data for $Ra$ covering seven decades, from
moderate ($10^8$) in DNS to very high ($10^{15}$) in Gottingen
experiments \cite{Ahlers2012a}. The agreement is very satisfactory.

Note that the theory, initially developed for horizontally
averaged MTP, is validated as shown in Fig.~\ref{fig:ltheta}. The agreement
in Fig.~\ref{fig:Theta} shows that it is also valid for MTP at a fixed location
near the sidewall ($x_p=0.0045$), with the same $z_{sub}^ +$ and slightly
different $z_{buf}^ +$ and $\rho$. Analysis
of MTPs at different $x$ locations in our DNS data shows that
$z_{buf}^+$ and $\rho$ have sensitive dependence on $x$ that reveals
internal mean-field structure of the momentum and temperature, which will be
reported elsewhere. Note also that a
common $z_{sub}^ + \approx 0.375$ yields a good description of both our
DNS with rectangular geometry and the cylindrical one \cite{Ahlers2012a},
as expected for universal near-wall variations.

\begin{figure}
\includegraphics[width=7cm]{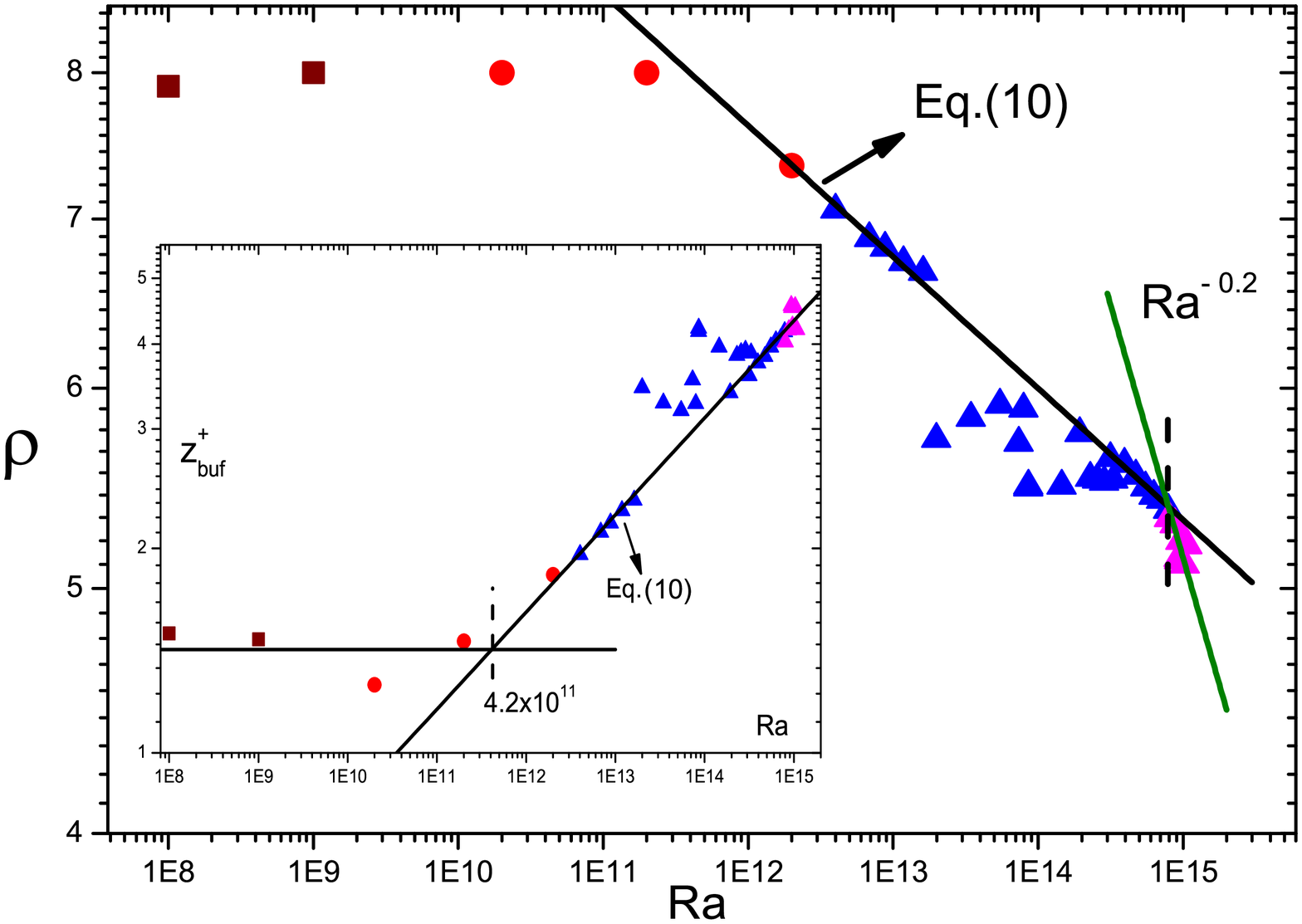}
\caption{(color). The variation of the two model parameters, $\rho$
(main) and $z^+_{buf}$ (inset), as a function of $Ra$, derived using
(\ref{log}) and (\ref{kapa}) from DNS data: brown squares (our
data), red discs \cite{Ahlers2012a}, blue and purple triangles
(experimental data \cite{Ahlers2012a}). Interpretations: for
$Ra<4.2\times 10^{11}$, the measuring point ($x=0.0045$) is
located within the sidewall kinetic boundary layer; for
$Ra>4.2\times 10^{11}$, the measuring point experiences strong
turbulence of the "classical regime", governed by (\ref{zbuf}); for
$Ra>8\times 10^{14}$, the ultimate regime \cite{He2012a} is
characterized by $ \rho  \propto Ra^{ -0.2}$.}
\label{fig:zbuf}
\end{figure}

Variation of the fitted $z_{buf}^+$ and $\rho$ with $Ra$ (Fig.~\ref{fig:zbuf})
reveals three transitions at $Ra_{c1}\approx 4\times
10^{11}$, $Ra_{c2}\approx 2\times 10^{13}$ and $Ra_{c3}\approx 8\times 10^{14}$.
The first plateau in Fig.~\ref{fig:zbuf} reveals the presence of weakly turbulent state close to the
sidewall at $x_p\approx 0.0045$ for $Ra<Ra_{c1}$. For $Ra>Ra_{c1}$,
$z_{buf}^+$ and $\rho$ begin to vary, indicating that strong fluctuations reach the probe.
In all cases, we observe a logarithmic MTP, but with no indication of a
kinetic logarithmic layer at moderate $Ra\sim Ra_{c1}$.
The second and the third transitions coincide with whose reported by \cite{Ahlers2012a},
indicating that the multi-layer parameters are sensitive to the variation of the physical states.
As shown in Fig.~\ref{fig:zbuf}, we find, for $Ra>Ra_{c1}$, $z_{buf}^ +$ and $\rho$
follow the power-law scaling:
\begin{equation}\label{zbuf}
z_{buf}^ + \approx 0.031Ra^{0.143},\quad \rho  \approx 32.1Ra^{ -
0.052}
\end{equation}
where the coefficients are obtained with a least-squares fitting.
The variation of $z_{buf}^+$ is mostly responsible for observed systematic variation of the slope
($A$), the inverse of which for shear flow is called the Karman
constant.

The scaling exponent close to $1/7$ in $z_{buf}^ +$
was predicted by Proccacia et al. \cite{Procaccia1991} based on a
earlier proposal of a mixing zone by the Chicago group
\cite{Heslot1987}, and experimentally detected by Xia et al.
\cite{SQZhou2002}. Thus, we speculate that an increasing buffer
layer thickness is a sign of increasing influence of the energetic
thermal plumes on the BL. Furthermore, we
consider $z_{buf}^ +$ to be a candidate measure of the height of the
plumes. This idea should be subjected to further tests by
experimental and DNS data. At this point, we caution about the universality of
the empirical laws (\ref{zbuf}) derived here from measurements at a
fixed distance from the sidewall \cite{Ahlers2012a}; variations of
$z_{buf}^+$ and $\rho$ need further study.

From (\ref{kapa}) and (\ref{zbuf}),
\begin{equation}
\kappa _\theta   = 1/A \approx 0.47Ra^{0.162}
\end{equation}
This predicts a significant variation of the log-law slope in RBC, as
compared to the universal Karman constant in the momentum transfer.
Our analysis shows that this variation is primarily due to the
change in $z_{buf}^ +$, interpreted above as the gradual increase of
the plume height with $Ra$. The last transition to the ultimate state
is characterized by a rapid decrease of $\rho$, leading to an
overall decrease of $\ell_\theta$ (at all distances from the wall).
We interpret this as a sign of plume correlation length crisis: for
$Ra > 8 \times 10^{14}$, at the ultimate state, motions become
significantly more turbulent, so that a stronger temperature
gradient is formed near the wall with a larger asymmetry of the bulk
temperature. The exact nature of this crisis
needs further investigation.

In summary, we offer the first analytic model for the temperature distribution
in the RBC, which is visibly superior to prior models giving temperature
variations only in the near-wall region (see \cite{Stevens2012,
Zhou2010}). The theory captures the documented small (logarithmic) variation and
slight asymmetry of the mean temperature in the bulk of the RBC cell, with a simple postulate of
the multi-layer scaling symmetry. Further interpretation
will be more intriguing: for instance, a new formula for $Nu$ can be
derived from the log-law for $\theta$:
\begin{equation}
Nu\approx A Ra^{1/7}\exp\{(Ra/Ra_c)^\beta\},
\end{equation}
which seems to yield a sensitive parameterization to interpret $Nu$-measurements
from different groups (to be reported). On the theoretical ground, further characterization of
detailed horizontal dependence would form a complete mean-field theory for the RBC system.
Finally, note that the successes reported
here and elsewhere \cite{She2012NJP,Zhang2012} support the view that the multi-layer picture is
universal to all wall-bounded turbulent flows.

We thank G. Ahlers and E. Bodenschatz for discussions and for sharing data.
This work is supported by National Nature Science Fund
90716008 and by MOST 973 project 2009CB724100.

\end{document}